\documentclass[preprint,5p,times,twocolumn]{elsarticle}

\usepackage{amssymb}
\usepackage{amsmath}
\usepackage{subcaption}
\usepackage[colorlinks=true, linkcolor=blue, citecolor=blue, urlcolor=blue, pdfauthor=author]{hyperref}
\usepackage[capitalise, nameinlink]{cleveref}
\biboptions{sort&compress}

\usepackage{lineno}

\usepackage{fancyhdr}
\journal{}
\date{}

\begin{document}

\begin{frontmatter}

\title{An accurate measurement of parametric array using a spurious sound filter topologically equivalent to a half-wavelength resonator}

\author[1]{Woongji Kim\fnref{equal}}
\ead{wj.kim@postech.ac.kr}
\author[1]{Beomseok Oh\fnref{equal}}
\ead{bs.oh@postech.ac.kr}
\author[1,2,3,4]{Junsuk Rho\corref{cor1}}
\ead{jsrho@postech.ac.kr}
\author[1]{Wonkyu Moon\corref{cor1}}
\ead{wkmoon@postech.ac.kr}
\cortext[cor1]{Corresponding authors}
\affiliation[1]{organization={Department of Mechanical Engineering, Pohang University of Science and Technology},
            addressline={77 Cheongam-ro, Nam-gu}, 
            city={Pohang},
            postcode={37673}, 
            state={Gyeongsangbuk},
            country={Republic of Korea}}
\affiliation[2]{organization={Department of Chemical Engineering, Pohang University of Science and Technology},
            addressline={77 Cheongam-ro, Nam-gu}, 
            city={Pohang},
            postcode={37673}, 
            state={Gyeongsangbuk},
            country={Republic of Korea}}
\affiliation[3]{organization={Department of Electrical Engineering, Pohang University of Science and Technology},
            addressline={77 Cheongam-ro, Nam-gu}, 
            city={Pohang},
            postcode={37673}, 
            state={Gyeongsangbuk},
            country={Republic of Korea}}
\affiliation[4]{organization={POSCO-POSTECH-RIST Convergence Research Center for Flat Optics and Metaphotonics},
            addressline={77 Cheongam-ro, Nam-gu}, 
            city={Pohang},
            postcode={37673}, 
            state={Gyeongsangbuk},
            country={Republic of Korea}}
\fntext[equal]{These authors contributed equally to this work.}

\begin{abstract}
Parametric arrays (PA) offer exceptional directivity and compactness compared to conventional loudspeakers, facilitating various acoustic applications. However, accurate measurement of audio signals generated by PA remains challenging due to spurious ultrasonic sounds arising from microphone nonlinearities. Existing filtering methods, including Helmholtz resonators, phononic crystals, polymer films, and grazing incidence techniques, exhibit practical constraints such as size limitations, fabrication complexity, or insufficient attenuation. To address these issues, we propose and demonstrate a novel acoustic filter based on the design of a half-wavelength resonator. The developed filter exploits the nodal plane in acoustic pressure distribution, effectively minimizing microphone exposure to targeted ultrasonic frequencies. Fabrication via stereolithography (SLA) 3D printing ensures high dimensional accuracy, which is crucial for high-frequency acoustic filters. Finite element method (FEM) simulations guided filter optimization for suppression frequencies at 40 kHz and 60 kHz, achieving high transmission loss (TL) around 60 dB. Experimental validations confirm the filter's superior performance in significantly reducing spurious acoustic signals, as reflected in frequency response, beam pattern, and propagation curve measurements. The proposed filter ensures stable and precise acoustic characterization, independent of measurement distances and incidence angles. This new approach not only improves measurement accuracy but also enhances reliability and reproducibility in parametric array research and development.

\end{abstract}




\end{frontmatter}

\thispagestyle{fancy}


\section{Introduction}
\label{sec:1}

Parametric acoustic arrays, first theorized by Westervelt \cite{Westervelt1963}, exploit the nonlinear interaction of intense ultrasonic waves to produce highly directional audible sound beams. Owing to their exceptional acoustic directivity and compactness compared to conventional acoustic sources, parametric arrays (PAs) have found applications in directional audio systems, personal audio communication, and active noise control \cite{Yoneyama1983, Gan2012, Zhong2022b}. Despite their versatility, the practical measurement and evaluation of PAs, particularly in the near-field region \cite{Zhong2022, Zhong2025}, remain technically challenging due to the inherent presence of spurious sound signals. These undesired signals primarily arise from the nonlinear responses of microphones and associated electronics when exposed to intense primary ultrasonic waves \cite{Bennett1975, Abuelmaatti2003, Ju2010, Ji2019}. The spurious sound, defined as the erroneous difference-frequency component measured by a microphone that does not correspond to the true acoustic signal in the air, significantly affects measurement accuracy \cite{Abuelmaatti2003}, particularly in the near-field region where primary ultrasonic wave amplitudes are elevated \cite{Zhong2022}.
\par
Several strategies have been proposed to mitigate the influence of spurious sounds in PA measurements, as comprehensively reviewed in previous literature \cite{Ji2016, Zhong2024}. Orientation of the microphone, leveraging its angular sensitivity characteristics, has been widely adopted as a simple approach. However, the efficacy of this method is significantly limited due to the modest transmission loss (TL) achievable at grazing incidence angles (e.g., 90\textdegree), which is insufficient for effectively suppressing intense ultrasonic signals and thus compromising measurement accuracy \cite{Ju2010, Nomura2022}. Physical acoustic filters placed in front of microphones represent another common approach. Early experiments by Bennett and Blackstock \cite{Bennett1975} employed dome-shaped acoustic filters, which significantly reduced ultrasonic signals while maintaining relatively low attenuation at audible frequencies. Subsequent studies utilized various materials, such as polymer films, aluminum plates, and phononic crystals, achieving considerable insertion losses at ultrasound frequencies \cite{Toda2005, Ye2011, Ji2016}. For example, Ye et al. implemented an aluminum plate filter providing more than 15 dB attenuation at ultrasonic frequencies above 30 kHz, yet less than 5 dB attenuation below 10 kHz \cite{Ye2011}. However, conventional filters often suffered limitations such as physical bulkiness, complex construction, or acoustic interference, significantly altering the measured acoustic fields, especially in the near field. The phase-cancellation technique offers an alternative solution by employing ultrasonic transducers excited with opposite phases to cancel the spurious signals along specific measurement axes. Despite its effectiveness, this approach is inherently limited in its spatial applicability \cite{Ji2012}. An optical interferometric method, utilizing interferometric detection of acoustic pressure without any mechanical components, has also been demonstrated as a promising spurious-sound-free measurement technique. Although optical interferometry offers contactless measurement capability and completely eliminates the spurious signal issue, it requires expensive, sophisticated equipment and complex numerical reconstruction, restricting its practical accessibility \cite{Ishikawa2021}.
\par
In this work, we propose a type of acoustic filter that effectively suppresses spurious sounds in PA measurements, thereby enhancing measurement accuracy and reliability to overcome the aforementioned limitations. Unlike conventional filters, the proposed filter excels in high-frequency suppression capability, compact size, minimal acoustic field perturbation, and notably, angular insensitivity. These advantages render the filter exceptionally well-suited to the growing interest in complex near-field measurements of parametric arrays exploiting local nonlinearities \cite{Zhong2022, Zhong2025}, significantly enhancing its practical applicability in such challenging acoustic environments. Through numerical and experimental validation, the present work aims to quantify the proposed filter's effectiveness, analyze its influence on the measured acoustic fields, and verify its applicability under diverse incidence angles and acoustic conditions.

\section{Current Techniques and Proposed Alternative}
\label{sec:2}

Several acoustic filtering strategies have been investigated to suppress unwanted spurious sounds induced during the measurement of PAs \cite{Bennett1975, Kamakura1984, Toda2005, Ji2016, Ju2010, Nomura2022}. Each of these methods, however, exhibits inherent limitations from the practical perspective, particularly in near-field environments or high-frequency ultrasound.

\subsection{Plastic Thin Film-based Acoustic Filters}
\label{subsec:21}

Plastic thin films have been utilized as acoustic filters to attenuate primary ultrasonic components in PA measurements. These filters function by exploiting the impedance mismatch between the thin film and the surrounding air, resulting in selective attenuation of high-frequency components while preserving the difference-frequency signal. Previous studies have demonstrated that such filters can provide attenuation levels of approximately 20--30 dB at ultrasonic frequencies, depending on the material properties and layer configuration \cite{Bennett1975, Kamakura1984, Toda2005}. In particular, multilayer polymer film arrangements have been employed to enhance attenuation performance within the 30--40 kHz range. Despite their potential, plastic thin film filters exhibit several practical limitations. The attenuation characteristics are highly sensitive to the spacing between layers, necessitating precise fabrication and alignment to achieve optimal performance. Additionally, single-layer films provide only moderate attenuation, requiring multi-layered configurations that introduce complexity in fabrication and handling. These factors limit the practical applicability of plastic thin film filters in PA measurements.

\subsection{Helmholtz Resonators}
\label{subsec:22}

Helmholtz resonators (HRs) possess promising filtering capabilities for targeted acoustic frequencies. However, at ultrasonic frequencies typically around 40 kHz and 60 kHz encountered in PAs, the physical dimensions required become exceedingly small. Specifically, cavity volumes and neck dimensions approach sub-millimeter scales, rendering precise fabrication extremely challenging or practically infeasible. Moreover, at these scales, even minor manufacturing deviations significantly alter the designed resonance conditions. Additionally, viscous and thermal losses become increasingly pronounced as dimensions shrink, reducing the quality factor (Q-factor) and diminishing the filter's selectivity. Consequently, despite the theoretical advantages of HRs, practical implementation as acoustic notch filters at ultrasonic frequencies around 40--60 kHz remains severely limited by manufacturing constraints and physical realities.

\subsection{Phononic Crystals}
\label{subsec:23}

Phononic crystals have attracted significant interest due to their unique capability to form band gaps through Bragg scattering. By arranging periodic structures (unit cells or scatterers) at distances satisfying the Bragg condition ($a=c/2f$), phononic crystals effectively serve as frequency-selective acoustic barriers \cite{Martinez-Sala1995, Lu2009}. However, this methodology presents two significant limitations. First, creating band gaps at relatively lower ultrasonic frequencies inherently requires a larger lattice constant, which leads to an increased overall filter size. For instance, lattice constants at 40 kHz and 60 kHz are approximately 4.3 mm and 2.9 mm, respectively. This requirement poses practical challenges for filter miniaturization and restricts the achievable number of unit cells within a limited spatial footprint, subsequently limiting performance enhancement. Second, phononic crystals fundamentally depend on planar wave incidence, causing a significant deterioration in performance under oblique incident angles. Consequently, their effectiveness is substantially reduced in complex near-field acoustic environments characterized by diverse incident wave directions.
\par
Moreover, the attenuation effect within the Bragg scattering bandgap is relatively weak, as seen in comparison to the local resonance bandgap. This is associated with the complex wavenumber in the band structure, which influences wave attenuation, and has been extensively explored in studies on phononic and photonic crystals \cite{Ao2009, Croenne2011}. Indeed, Ref. \cite{Ji2016} reported a moderate TL of around 30 dB, which is insufficient for accurate measurements in high-intensity ultrasound. Therefore, the use of phononic crystals is not suitable for applications requiring strong attenuation.

\subsection{Grazing Incidence of Microphones}
\label{subsec:24}

Adjusting the angle of incidence of microphones (e.g., placing microphones at grazing angles) provides a straightforward technique for spurious sound reduction by exploiting microphone directivity at ultrasonic frequencies \cite{Ju2010, Nomura2022}. This method offers significant practical advantages as it requires no specialized equipment or additional acoustic components, enabling direct and uncomplicated application of measurement microphones. Moreover, its simplicity makes it particularly useful in near-field measurements, where space limitations favor this method. However, the limited attenuation achievable with this approach restricts its effectiveness, making it less suitable for spurious sound suppression in such conditions. The achievable attenuation using this method is modest, typically limited to approximately 20 dB at best. In scenarios characterized by high-intensity ultrasonic fields, such as near-field measurements, this limited attenuation proves insufficient for effectively reducing intermodulation distortion (IMD) \cite{Nomura2022}. Furthermore, because this method does not allow explicit control over attenuation characteristics specifically tailored to the source frequency spectrum, its practical efficacy remains restricted compared to dedicated acoustic filtering solutions.

\subsection{Motivation for the Proposed Half-Wavelength Acoustic Filter}
\label{subsec:25}

Considering the aforementioned limitations, there remains a long-standing demand for a compact, easily manufacturable acoustic filter that can effectively suppress spurious sound components without significant dependence on wave incidence angles. To address these requirements, this study presents a half-wavelength resonator-based filter that effectively reduces spurious sound, ensuring that the measured acoustic pressure accurately reflects the true signal from a PA by reducing IMD-induced distortions. Furthermore, drawing on the concept of using a single, simple-shaped homogeneous element\textemdash much simpler than the periodic structures found in phononic crystal filters \cite{Cervenka2016}\textemdash\ the proposed design can be readily fabricated via widely available 3D printing technologies, thereby overcoming practical manufacturing challenges. In addition, unlike phononic crystal filters, this filter demonstrates robust attenuation performance regardless of the incident angle of ultrasonic waves, making it exceptionally suitable for accurate acoustic measurements in complex near-field induced PA phenomena.

\section{Concept and Design of Proposed Spurious Sound Filter}
\label{sec:3}

\subsection{Operational Principle of Half-Wavelength Resonator}
\label{subsec:31}

\begin{figure}[ht]
    \centering
    \begin{subfigure}[t]{4.25cm}
      \centering
      \includegraphics[width=\textwidth]{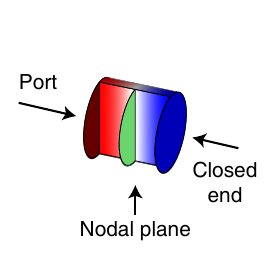}
      \caption{}
      \label{fig:1a}
    \end{subfigure}
    \hfill
    \begin{subfigure}[t]{4.25cm}
      \centering
      \includegraphics[width=\textwidth]{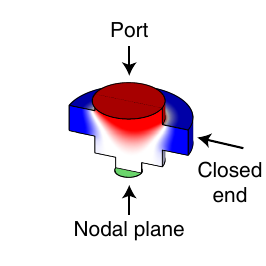}
      \caption{}
      \label{fig:1b}
    \end{subfigure}
    \\
    \begin{subfigure}[b]{4.25cm}
      \centering
      \includegraphics[width=\textwidth]{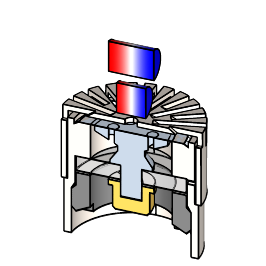}
      \caption{}
      \label{fig:1c}
    \end{subfigure}
    \hfill
    \begin{subfigure}[b]{4.25cm}
      \centering
      \includegraphics[width=\textwidth]{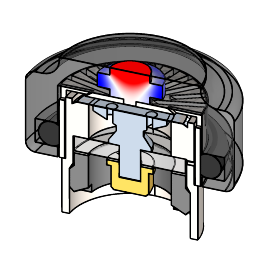}
      \caption{}
      \label{fig:1d}
    \end{subfigure}
    
    \caption{(a) Schematic representation of a half-wavelength closed-end pipe resonator. (b) Modified half-wavelength resonator geometry designed to shift the nodal plane outward for practical measurement applications. (c) Conventional half-wavelength resonator geometry integrated with a standard measurement microphone, illustrating the inherent limitation at high frequencies. (d) Proposed modified resonator geometry integrated with a microphone, effectively relocating the nodal plane to an accessible external position, enabling practical high-frequency acoustic pressure measurements. (Red regions represent the acoustic pressure inlet (port), blue regions indicate acoustically closed ends, and green region denote the nodal plane.)}
    \label{fig:1}
\end{figure}

The proposed acoustic filter operates based on the acoustic resonance phenomenon occurring within a half-wavelength resonator pipe with one end acoustically closed. As illustrated in \cref{fig:1a}, when acoustic waves enter from the open port, a standing wave is formed inside the resonator at a frequency determined by the pipe length. Specifically, if the pipe length is exactly half the wavelength of the targeted frequency, a distinctive pressure distribution emerges: a nodal plane (pressure node) naturally forms at the midpoint of the pipe. At this nodal plane, the acoustic pressure theoretically approaches zero, effectively resulting in maximum TL at that frequency. Thus, placing a measurement microphone precisely at this nodal plane enables significant suppression of the corresponding acoustic frequency, greatly minimizing the measured acoustic pressure.
\par
However, the practical implementation of this structure is challenging at higher frequencies due to shorter wavelengths, resulting in resonator dimensions significantly smaller than typical microphone dimensions. This limitation is illustrated clearly in \cref{fig:1c}, where the conventional resonator structure becomes impractical, making accurate pressure measurement at the internal nodal plane difficult or impossible. To overcome this issue, the resonator geometry should be modified, as shown in \cref{fig:1b}. Consequently, the modified design can be implemented in practice using the measurement microphone depicted in \cref{fig:1d}. This reshaped resonator preserves fundamental resonant characteristics while relocating the nodal plane outward as an external surface, ensuring direct and convenient placement of the microphone for accurate acoustic measurements.
\par
These two configurations—conventional (\cref{fig:1a,fig:1c}) and modified (\cref{fig:1b,fig:1d})—are topologically equivalent, sharing identical acoustic characteristics despite their differing physical forms. This topological transformation thus maintains the essential acoustic filtering principle, greatly enhancing practical applicability and enabling precise measurement even at high-frequency acoustic scenarios.

\subsection{Design Methodology}
\label{subsec32}

\begin{figure}[t]
    \centering
    \includegraphics[width=8.5cm]{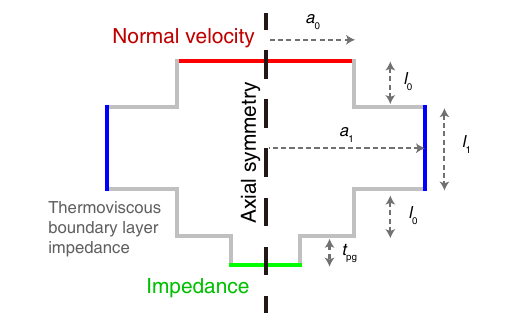}
    \caption{Cross-sectional schematic of the proposed acoustic filter illustrating its simplified geometry. The red boundary at the top represents the acoustic input port with a normal velocity boundary condition. The blue boundary denotes acoustically closed ends with thermal-viscous boundary layer impedance. The green boundary at the bottom indicates the nodal plane, where the measurement microphone is positioned.}
    \label{fig:2}
\end{figure}

The proposed acoustic filter features a simple structure, as illustrated in \cref{fig:2}. This simplicity reduces the number of required design parameters, facilitating straightforward optimization, and enables precise fabrication via SLA 3D printing, particularly advantageous when working at submillimeter scales. The principal design parameters include the primary duct radius $a_0$ and its length $l_0$, where $l_0$ also represents the minimum achievable wall thickness in the 3D printing process. The expanded duct length, $l_1$, is also a design parameter, while the expanded duct radius, $a_1$, serves as sole design variables in the optimization process.
\par
The dimensions of the acoustic duct were conceptually established based on the targeted frequency for suppression, $f_0$. For a 60 kHz, the half-wavelength is approximately 2.9 mm. The input port diameter was set to a comparable dimension of 3 mm. This choice ensures that the duct dimension exceeds the quarter-wavelength limit, promoting the formation of a two-dimensional cylindrical wave distribution instead of a one-dimensional wave along the duct axis. Consequently, this structure relocates the nodal plane externally, making it practical for microphone placement and accurate pressure measurement. Furthermore, restricting the acoustic wave path exclusively through the microphone protection grid's center hole simplifies boundary condition definitions for modeling purposes. Specifically, the Brüel \& Kjær Type 4192 measurement microphone features a protection grid with a center hole diameter of 1.15 mm, while the grid's acoustic slit wedge starts from a diameter of approximately 3.8 mm. Setting the input port diameter to 3 mm thus ensures a direct and exclusive acoustic connection through the center hole.
\par
In the finite element model, a normal velocity boundary condition was applied to the acoustic input boundary, indicated in red. At the targeted suppression frequency, $f_0$, the boundary condition on the nodal plane (marked in green)\textemdash the surface at which the microphone diaphragm measures acoustic pressure\textemdash should be modeled as an impedance boundary condition based on the microphone diaphragm's acoustic impedance and calibrator load volume data provided by the manufacturer \cite{bksv2019}. Here, the calibrator load volume refers to the combined volume consisting of the effective acoustic volume associated with diaphragm compliance and the air trapped between the diaphragm and the interior surfaces of the microphone protection grid \cite{bksv1995}. However, since acoustic wave propagation to the diaphragm occurs solely through the center hole of the protection grid, the volume occupied by this hole was subtracted from the calibrator load volume. Thus, the acoustic impedance boundary condition at the microphone interface (nodal plane) could be expressed as

\begin{equation}
Z=\frac{V_{\mathrm{lf}}}{V_{\mathrm{cal}}-V_{\mathrm{ch}}} \left(j\omega L_{\mathrm{ds}} + R_{\mathrm{ds}} + \frac{1}{j \omega C_{\mathrm{ds}}}\right)
\end{equation}

\noindent where $V_{\mathrm{lf}}$ represents the low-frequency volume, while $C_{\mathrm{ds}}$, $L_{\mathrm{ds}}$, and $R_{\mathrm{ds}}$ correspond to the acoustic compliance, mass, and damping resistance of the diaphragm system, respectively. $V_{\mathrm{cal}}$ denotes the calibrator load volume. Additionally, $V_{\mathrm{ch}}$ represents the volume of the protection grid's center hole, which is calculated as $\pi d_{\mathrm{ch}}^2/4 \times t_{\mathrm{pg}}$, where $d_{\mathrm{ch}}$ is the diameter of the center hole and $t_{\mathrm{pg}}$ is the thickness of the protection grid. Additionally, given the narrow geometry of the acoustic duct, thermal and viscous boundary layer impedances were applied to the duct walls (colored gray and blue), ensuring accurate representation of acoustic losses. The model was solved as an axisymmetric 2D simulation using COMSOL Multiphysics with the pressure acoustics-frequency domain module. Optimization of the duct dimensions was conducted using the BOBYQA method, aiming to maximize TL at the desired suppression frequency, $f_0$, with the expanded duct radius $a_1$ as the design variable, for a fixed duct length $l_1$. The optimization process was terminated upon reaching an optimality tolerance of $1\times10^{-5}$. The geometry of a resonant filter element in a variable-cross-section waveguide has a profound impact on its transmission coefficient \cite{Cervenka2016}.

\begin{figure}[t]
  \centering
  \begin{subfigure}[t]{8.5cm}
    \centering
    \includegraphics[width=\textwidth]{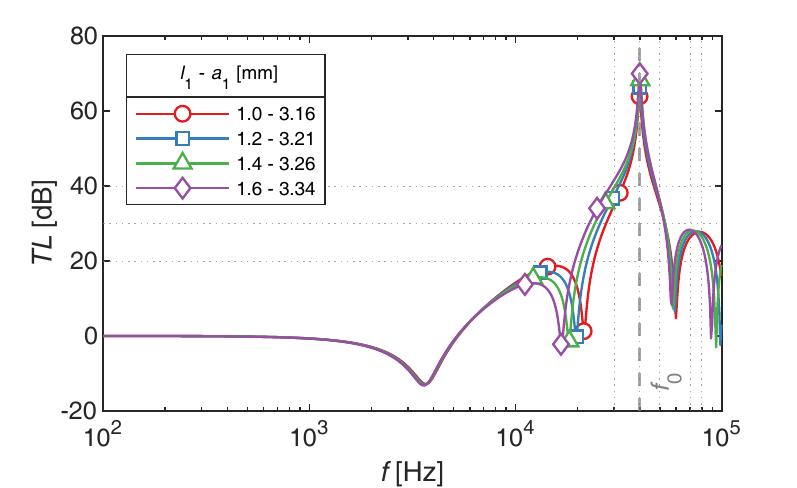}
    \caption{}
    \label{fig:3a}
  \end{subfigure}
  \begin{subfigure}[t]{8.5cm}
    \centering
    \includegraphics[width=\textwidth]{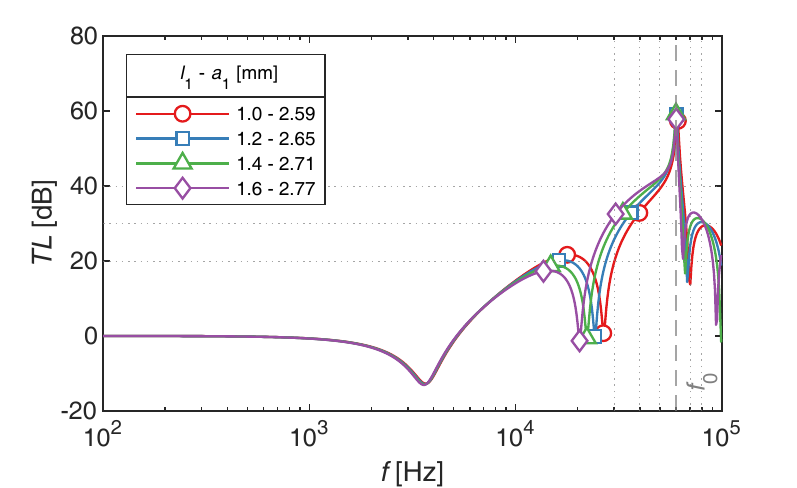}
    \caption{}
    \label{fig:3b}
  \end{subfigure}
  \caption{TL characteristics of the acoustic filters optimized for suppression frequencies of (a) 40 kHz and (b) 60 kHz, respectively. Each curve corresponds to different pairs of fixed duct lengths, $l_1$, and their optimized duct radii, $a_1$.}
  \label{fig:3}
\end{figure}

\subsection{Practical Design Examples}
\label{subsec33}

\Cref{fig:3a} illustrates the TL characteristics obtained through FEM optimization of the proposed acoustic filter for spurious sound suppression at a target frequency of 40 kHz. Optimization was performed by pairing fixed values of the design parameter, $l_1$, with their corresponding optimized values of the design variable, $a_1$. Due to the underlying resonant principle, extremely high TLs around 60 dB were achieved precisely at the desired suppression frequency, $f_0$. Furthermore, the bandwidth over which TL remains above 40 dB extends approximately 10 kHz, indicating a highly effective filtering capability within the typical audio frequency range generated by PAs. The microphone's intermodulation distortion (IMD), which causes spurious sounds, increases when the difference between the carrier and sideband frequencies is small\textemdash that is, at low audio or difference frequencies. Consequently, even a filter with a 10 kHz bandwidth can effectively eliminate strong spurious sounds, making it sufficient for practical applications. Similarly, \Cref{fig:3b} presents optimized results for the acoustic filter designed for a suppression frequency of 60 kHz, showing nearly identical performance trends. This consistent TL response characteristic demonstrates that the proposed principle and optimization method allow for the design of filters tailored to different suppression frequencies, illustrating that suppression can be achieved at various target frequencies by systematically adjusting the design parameters and optimization variables.
\par
Notably, both designs consistently exhibit a TL dip around 3 kHz within the audio frequency band, which is attributed to the microphone's characteristic. Since this behavior is predictable and reproducible, it can be effectively compensated during practical implementations by measuring the actual TL prior to experiments and applying appropriate corrections. Thus, the presence of this low-frequency resonance is unlikely to result in significant operational challenges.

\section{Fabrication and Experimental Validation}
\label{sec:4}

\subsection{Fabrication using 3D Printing Technology}
\label{subsec:41}

\begin{figure}[t]
    \centering
    \includegraphics[width=8.5cm]{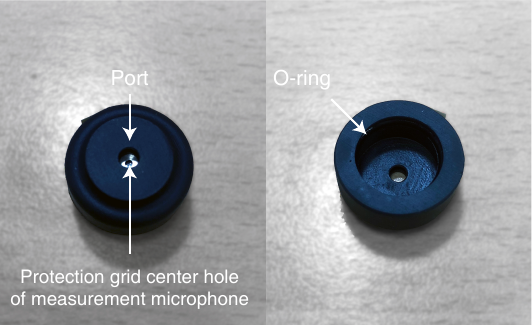}
    \caption{Photograph of the fabricated acoustic filter using SLA 3D printing. (Left) The top view shows the port aligned with the protection grid center hole of the microphone, allowing controlled acoustic transmission. (Right) The bottom view displays the O-ring, which is seated in the O-ring groove, ensuring a secure seal when the microphone is inserted into the designated mounting space.}
    \label{fig:4}
\end{figure}

The proposed acoustic filter was fabricated using SLA 3D printing technology, which was selected for its high precision and ability to produce fine structural details critical to achieving the resonator's intended frequency response in the design. The SLA 3D printing process offers dimensional accuracy within tens of micrometers, ensuring that the designed acoustic filter can be manufactured with high fidelity \cite{Nulty2022}.
The dimensional accuracy of the fabricated duct geometry is particularly important, as small variations in the design parameters directly influence the filter's acoustic performance. For example, when the extended duct length, $l_1$, changes from 1.2 to 1.4 mm of the 40 kHz filter, the corresponding optimized duct radius, $a_1$, changes from 3.21 to 3.26 mm. Since the resolution of SLA 3D printing is well within this range, the printer can reliably produce these dimensional variations, ensuring the filter functions as intended. This precision level ensures that the TL characteristics predicted during optimization are accurately realized in the fabricated filter. \Cref{fig:4} shows the filter fabricated using SLA 3D printing, where both the 40 kHz and 60 kHz designs have an $l_1$ of 1.4 mm.

\subsection{Experimental Setup and Methodology}
\label{subsec:42}

\begin{figure}[t!]
  \centering
  \begin{subfigure}[t]{8.5cm}
    \centering
    \includegraphics[width=\textwidth]{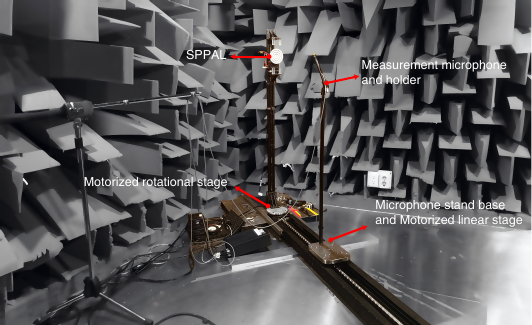}
    \caption{}
    \label{fig:5a}
  \end{subfigure}
  \begin{subfigure}[t]{8.5cm}
    \centering
    \includegraphics[width=\textwidth]{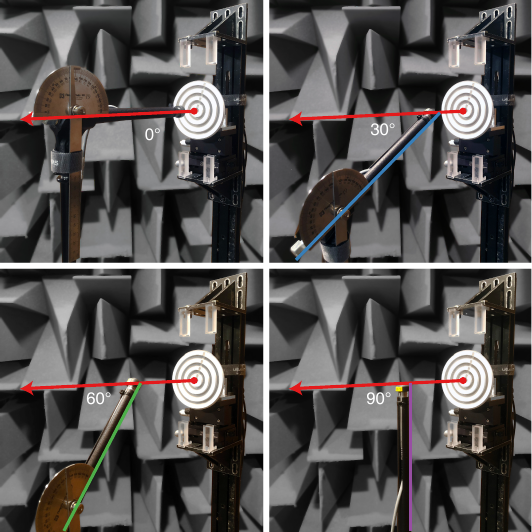}
    \caption{}
    \label{fig:5b}
  \end{subfigure}

  \caption{Experimental setup for measuring the TL of the acoustic filter. (a) Overview of the measurement system inside the semi-anechoic chamber, consisting of a SPPAL as the acoustic source, a measurement microphone with a holder mounted on a stand, and a motorized system comprising a rotational stage and a linear stage for precise positioning. (b) Adjustment of the microphone holder to control the angle of incidence ($\theta_i$), set at 0\textdegree, 30\textdegree, 60\textdegree, and 90\textdegree, respectively, to evaluate the filter's response across different incidence angles.}
  \label{fig:5}
\end{figure}

Acoustic measurements were conducted in a semi-anechoic chamber at Pohang University of Science and Technology (POSTECH). The chamber had a background noise level below 30 dB SPL, a free-field effective volume of $3\times3\times2$ $ \textrm{m}^3$, an estimated absorption coefficient near 0.99, and a low cutoff frequency of 150 Hz. During measurements, environmental conditions were maintained at a temperature of 20\textcelsius\ and a relative humidity of 30 \%. A modified version of stepped plate parametric array loudspeaker (SPPAL) \cite{Oh2023} was mounted on a motorized rotational stage using a custom-made jig, and positioned at 1.4 m above the chamber floor, used lower sideband amplitude modulation (LSB-AM) to generate audio signals. The audio sound acoustic field measurements were performed using a Brüel \& Kjær 1/2-inch microphone (Type 4192), both with and without the acoustic filter installed. The angle of incidence was adjusted by controlling the relative orientation between the acoustic source and the measurement microphone, as illustrated in \cref{fig:5b}. This was achieved by rotating the microphone holder to set the acoustic axis at incidence angles of 0\textdegree, 30\textdegree, 60\textdegree, and 90\textdegree, allowing an angular robustness of the filter's performance across different incident angles. A dynamic signal analyzer (SR785, Stanford Research Systems) was utilized to generate the excitation signals driving the transducer through a power amplifier (HSA4052, NF Corporation), and to simultaneously acquire the microphone output signals conditioned via a conditioning amplifier (Type 2690, Brüel \& Kjær). An electrical filter with a passband from 20 Hz to 100 kHz was incorporated to prevent electrical attenuation from influencing the measured results. MATLAB software was employed for automated data acquisition and instrument control.

\subsection{Measurement Results}
\label{subsec:43}

\subsubsection{Transmission loss of acoustic filter}
\label{subsubsec:431}

\begin{figure*}[t!]
  \centering
  \begin{subfigure}[t]{8.5cm}
    \centering
    \includegraphics[width=\textwidth]{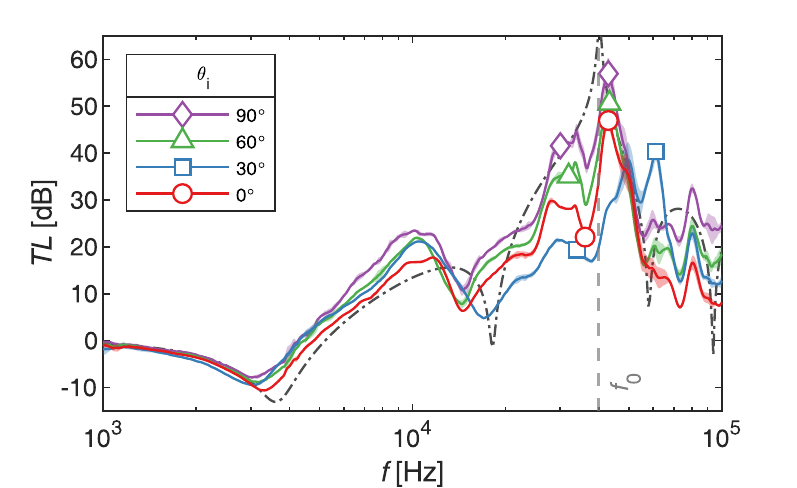}
    \caption{}
    \label{fig:6a}
  \end{subfigure}
  \hfill
  \begin{subfigure}[t]{8.5cm}
    \centering
    \includegraphics[width=\textwidth]{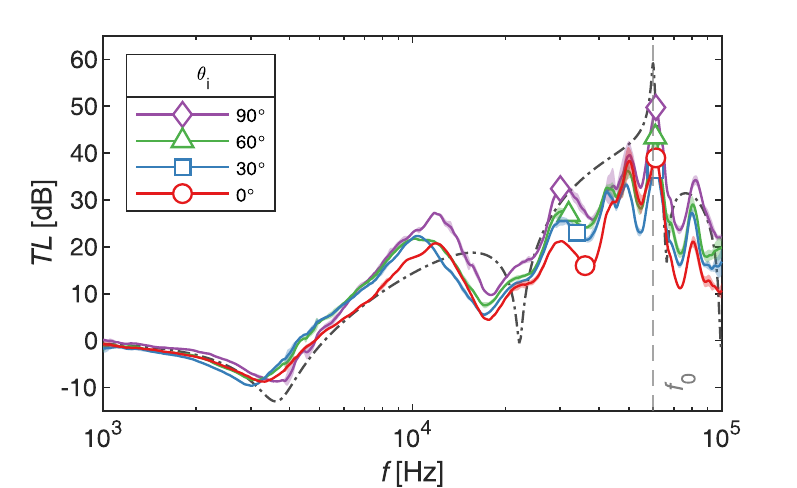}
    \caption{}
    \label{fig:6b}
  \end{subfigure}
  \\
  \begin{subfigure}[b]{8.5cm}
    \centering
    \includegraphics[width=\textwidth]{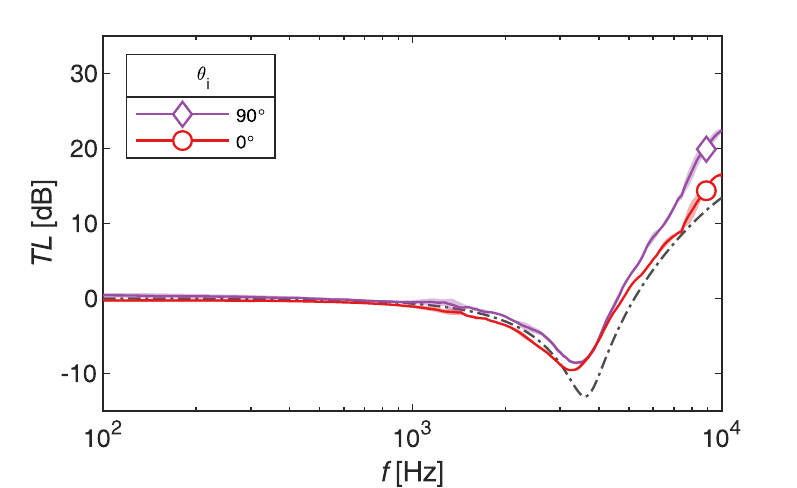}
    \caption{}
    \label{fig:6c}
  \end{subfigure}
  \hfill
  \begin{subfigure}[b]{8.5cm}
    \centering
    \includegraphics[width=\textwidth]{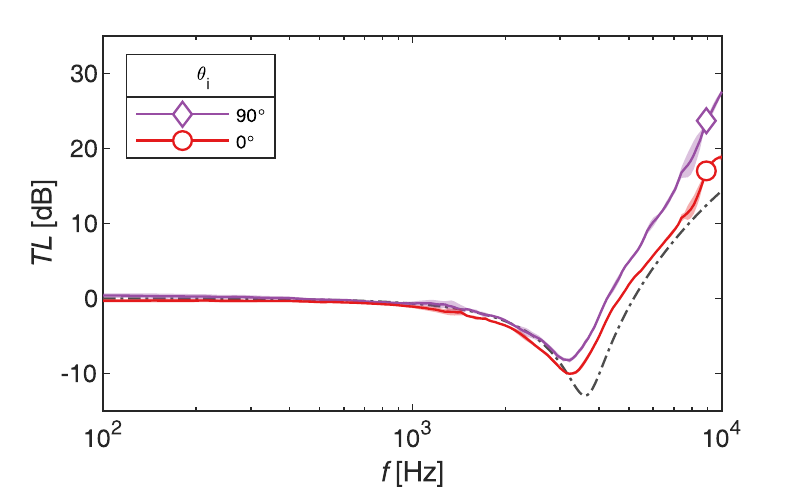}
    \caption{}
    \label{fig:6d}
  \end{subfigure}

  \caption{TL measurements of the fabricated acoustic filters for suppression frequencies of (a, c) 40 kHz and (b, d) 60 kHz. The black dash-dotted lines represent FEM simulations, while the vertical gray dashed lines indicate the targeted suppression frequencies, $f_0$. The red, blue, green, and purple solid lines correspond to incidence angles of 0\textdegree, 30\textdegree, 60\textdegree, and 90\textdegree, respectively. Each TL curve represents the average of measurements taken at distances of 0.1, 0.2, 0.5, and 1,0 m, with the small shaded regions illustrating the standard deviation across distances, demonstrating excellent consistency.}
  \label{fig:6}
\end{figure*}

TL quantifies the reduction in acoustic energy as a wave propagates through a system, typically expressed in decibels (dB) as a function of the power ratio. Since TL is inherently a power-based measure, while microphone measurements capture acoustic pressure, TL is derived from pressure measurements as follows:

\begin{equation}
  TL=20 \log_{10}\left|\frac{p_{\mathrm{without}}}{p_{\mathrm{with}}}\right|
\end{equation}

\noindent where $p_{\mathrm{without}}$ and $p_{\mathrm{with}}$ denote the measured sound pressures at the microphone position without and with the acoustic filter installed, respectively. To obtain TL values, measurements were first conducted with the microphone exposed directly to the incident sound field without the filter, followed by a second set of measurements with the filter in place. The difference in recorded pressure levels was then used to compute TL at each frequency of interest.
\par
\Cref{fig:6} shows the measured TL characteristics of the fabricated acoustic filters designed for suppression frequencies of 40 kHz and 60 kHz, respectively. The black dash-dotted lines represent the numerical simulation results obtained from FEM, and the vertical gray dashed lines indicate the targeted suppression frequencies, $f_0$. The solid-colored lines represent experimentally measured TL values, averaged from measurements taken at distances of 0.1, 0.2, 0.5, and 1 m from the source. The red, blue, green, and purple solid lines correspond to incidence angles of 0\textdegree, 30\textdegree, 60\textdegree, and 90\textdegree, respectively. The shaded regions around each line, representing the standard deviation across measurements at different distances, are nearly indistinguishable due to minimal discrepancies. This indicates that the measured TL remains consistent and largely independent of distance. Moreover, the increase in TL with incidence angle demonstrates that the proposed acoustic filter benefits not only from its inherent suppression mechanism but also from the grazing incidence method, further enhancing its overall performance.
\par
Overall, the experimental results closely follow the trends predicted by FEM. Although minor variations in TL can be observed with changing incidence angles, these differences remain relatively small, indicating that the proposed acoustic filter provides robust suppression across a wide angular range. Measurements were performed across a frequency range from approximately 1 kHz to 100 kHz, with higher-frequency measurements carried out using the SPPAL, a source optimized for ultrasonic frequency generation. Due to limitations in the low-frequency output of the SPPAL, the flat TL below 1 kHz was measured separately using a baffled loudspeaker. These low-frequency measurements serve as baseline references for calibrating and correcting the frequency response, beam patterns, and propagation characteristics of audio signals generated by the SPPAL in subsequent analyses.
\par
The slight discrepancies between the FEM results and experimental data could originate from manufacturing tolerances in the resonant filter dimensions. Moreover, the FEM analysis performed here did not include nonlinear acoustic phenomena, such as amplitude-dependent shifts in resonance frequencies or nonlinear harmonic interactions, which are known to occur in acoustic resonators with varying cross-sections at high acoustic amplitudes \cite{Cervenka2016, Hamilton2001}. Nevertheless, considering the overall close agreement between the numerical predictions and measurements, these minor deviations do not significantly compromise the reliability and applicability of the proposed acoustic filter. In conclusion, the developed acoustic filter achieves stable, distance-independent TL performance, exhibits a predictable directivity pattern with respect to incidence angles, and provides effective suppression of ultrasound at the targeted frequencies.

\subsubsection{Audio sound frequency response}
\label{subsubsec:432}

\begin{figure}[t]
  \centering
  \begin{subfigure}[t]{8.5cm}
    \centering
    \includegraphics[width=\textwidth]{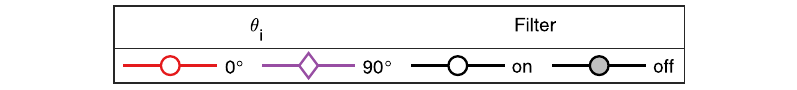}
  \end{subfigure}
  \\
  \begin{subfigure}[t]{4.25cm}
    \centering
    \includegraphics[width=\textwidth]{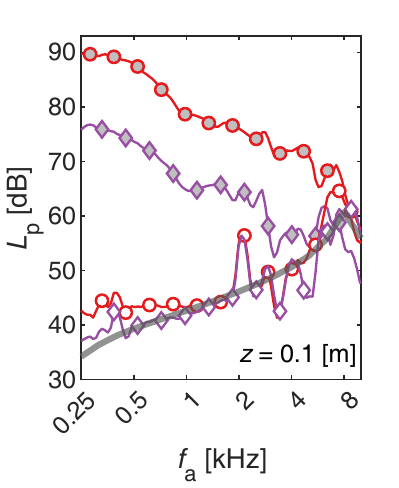}
    \caption{}
    \label{fig:7a}
  \end{subfigure}
  \hfill
  \begin{subfigure}[t]{4.25cm}
    \centering
    \includegraphics[width=\textwidth]{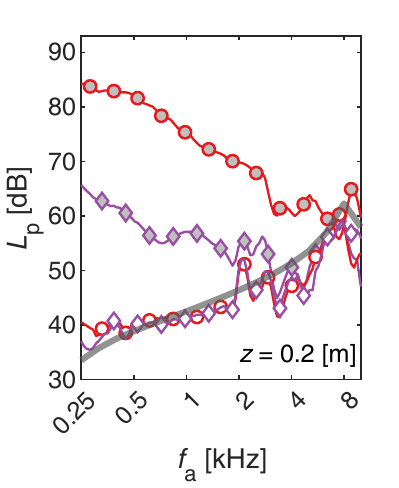}
    \caption{}
    \label{fig:7b}
  \end{subfigure}
  \\
  \begin{subfigure}[t]{4.25cm}
    \centering
    \includegraphics[width=\textwidth]{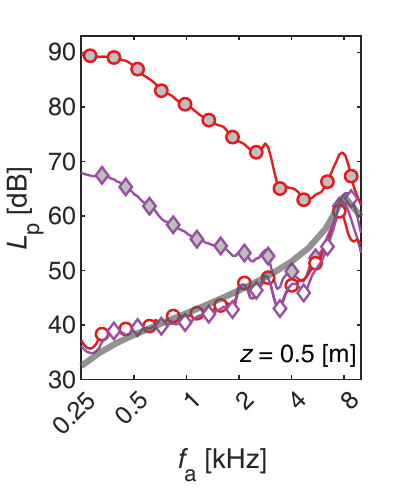}
    \caption{}
    \label{fig:7c}
  \end{subfigure}
  \hfill
  \begin{subfigure}[t]{4.25cm}
    \centering
    \includegraphics[width=\textwidth]{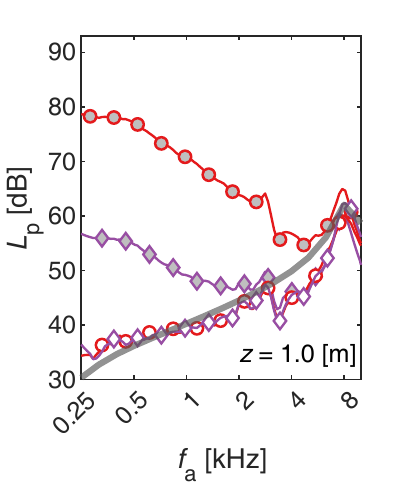}
    \caption{}
    \label{fig:7d}
  \end{subfigure}

  \caption{Measured frequency responses of the audio sound generated by the SPPAL at distances of (a) 0.1 m, (b) 0.2 m, (c) 0.5 m, and (d) 1.0 m from the acoustic source. Gray-filled markers represent measurements without the acoustic filter, while white-filled markers indicate measurements with the filter installed. The incidence angles of 0\textdegree\ (red circles) and 90\textdegree\ (purple diamonds) are distinguished by color and marker shape. The thick gray solid lines represent numerical simulation results.}
  \label{fig:7}
\end{figure}

\Cref{fig:7} shows the measured frequency responses of the audio sound generated by the SPPAL, evaluated at distances of 0.1, 0.2, 0.5, and 1.0 m from the acoustic source. The characteristics of the SPPAL and the nonlinear acoustic analysis methodology are briefly provided in \ref{app:1}, with the results of this nonlinear analysis depicted as thick gray solid lines. Red circles correspond to an incident angle of 0\textdegree, while purple diamonds denote an incident angle of 90\textdegree. The marker filled color indicates the filter condition: white-filled markers represent measurements with the filter installed, while gray-filled markers indicate measurements without the filter; these line and marker conventions are consistently applied in the subsequent figures. Measurements conducted without the acoustic filter demonstrate substantial levels of spurious sound, particularly at the 0\textdegree\ incidence angle, where undesired ultrasonic components significantly overshadow the true audio signal. Although grazing incidence measurements (90\textdegree) reduce the intensity of spurious components by approximately 20 dB, this reduction deviates significantly from the behavior observed in the low-frequency, indicating that the grazing incidence approach alone is inadequate for complete suppression. Additionally, this approach inherently lacks tunability to specifically account for the ultrasound frequency characteristics of the acoustic source (refer to \cref{fig:b1} for TL measurements using the grazing incidence method), unlike the proposed acoustic filter.
\par
In contrast, measurements with the acoustic filter installed show notable suppression of spurious sound, improving with increased measurement distance. At the shortest distance (0.1 m), residual spurious components persist at lower frequencies, especially at the 0\textdegree\ incidence angle. In this scenario, combining the acoustic filter with the grazing incidence method, which provides additional TL, enables accurate measurements even in extremely intense ultrasonic fields near the acoustic source. At greater distances (0.2, 0.5, and 1.0 m), the frequency responses at 0\textdegree\ and 90\textdegree\ incidence angles closely coincide, clearly demonstrating the proposed filter's effectiveness in mitigating spurious sounds across the audible frequency range.
\par
Numerical simulation results obtained from the spherical wave expansion (SWE) method closely align with experimental measurements, validating the accuracy of numerical predictions and confirming the practical efficacy of the fabricated acoustic filter.

\subsubsection{Audio sound beam pattern}
\label{subsubsec:433}

\begin{figure*}[t!]
  \centering
  \begin{subfigure}[t]{8.5cm}
    \centering
    \includegraphics[width=\textwidth]{Figure/FigureLegend.pdf}
  \end{subfigure}
  \\
  \begin{subfigure}[t]{4.25cm}
    \centering
    \includegraphics[width=\textwidth]{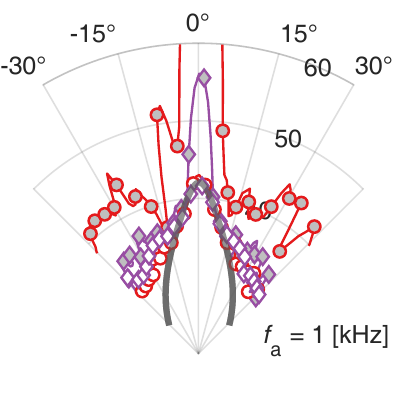}
    \caption{}
    \label{fig:8a}
  \end{subfigure}
  \hfill
  \begin{subfigure}[t]{4.25cm}
    \centering
    \includegraphics[width=\textwidth]{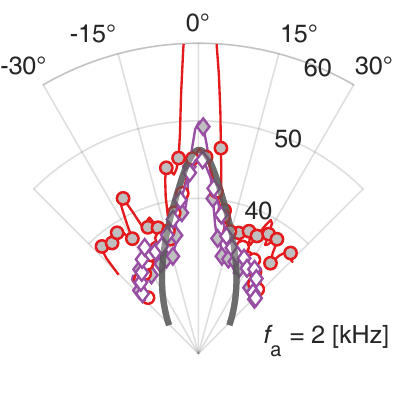}
    \caption{}
    \label{fig:8b}
  \end{subfigure}
  \hfill
  \begin{subfigure}[t]{4.25cm}
    \centering
    \includegraphics[width=\textwidth]{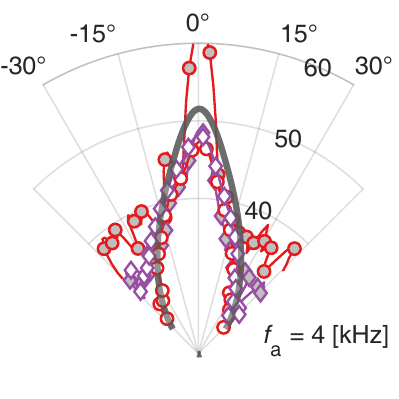}
    \caption{}
    \label{fig:8c}
  \end{subfigure}
  \hfill
  \begin{subfigure}[t]{4.25cm}
    \centering
    \includegraphics[width=\textwidth]{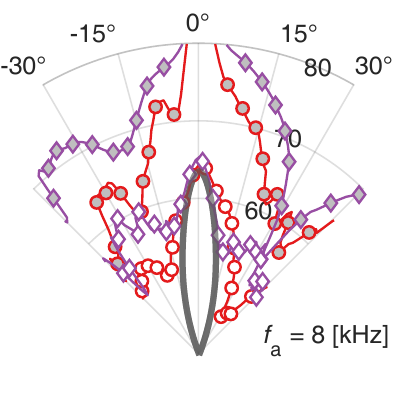}
    \caption{}
    \label{fig:8d}
  \end{subfigure}
  \\
  \begin{subfigure}[t]{4.25cm}
    \centering
    \includegraphics[width=\textwidth]{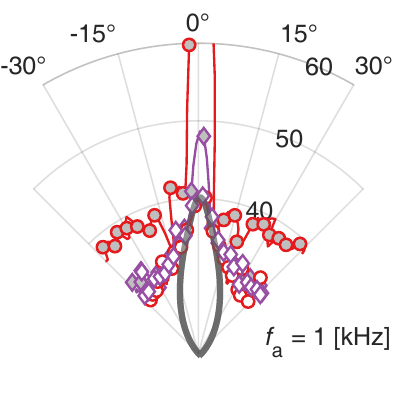}
    \caption{}
    \label{fig:8e}
  \end{subfigure}
  \hfill
  \begin{subfigure}[t]{4.25cm}
    \centering
    \includegraphics[width=\textwidth]{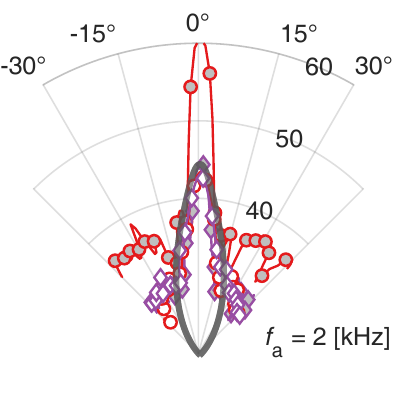}
    \caption{}
    \label{fig:8f}
  \end{subfigure}
  \hfill
  \begin{subfigure}[t]{4.25cm}
    \centering
    \includegraphics[width=\textwidth]{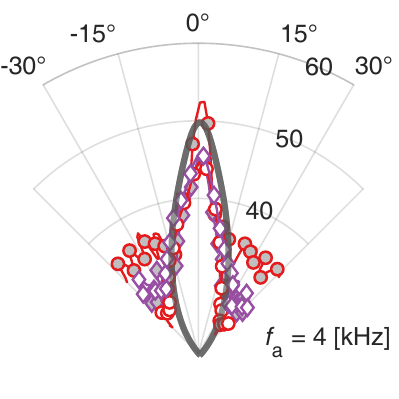}
    \caption{}
    \label{fig:8g}
  \end{subfigure}
  \hfill
  \begin{subfigure}[t]{4.25cm}
    \centering
    \includegraphics[width=\textwidth]{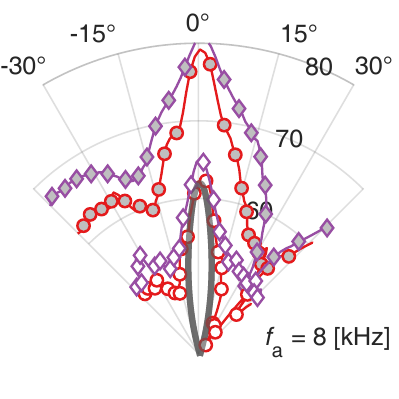}
    \caption{}
    \label{fig:8h}
  \end{subfigure}
  
  \caption{Measured unnormalized beam patterns of audio signals from the SPPAL at distances of 0.5 m (a\textendash d) and 1.0 m (e\textendash h) in dB SPL. Each subfigure corresponds to a specific audio frequency: (a, e) 1 kHz, (b, f) 2 kHz, (c, g) 4 kHz, and (d, h) 8 kHz. The line color and marker type indicate the incidence angle, where red circles represent 0\textdegree\ and purple diamonds denote 90\textdegree. The marker filled color indicates the filter condition: white-filled markers represent measurements with the filter installed, while gray-filled markers indicate measurements without the filter. The thick gray solid lines depict the beam patterns predicted by the SWE method, providing a reference for comparison with experimental data.}
  \label{fig:8}
\end{figure*}

\Cref{fig:8} presents the measured unnormalized beam patterns of audio signals generated by the SPPAL at distances of 0.5 m and 1.0 m from the acoustic source. Beam patterns were evaluated at representative audio frequencies of 1, 2, 4, and 8 kHz, as indicated by the legend. In the absence of the filter, significant spurious sound components lead to artificially exaggerated directivity, particularly at the 0\textdegree\ incidence angle, compared to measurements at 90\textdegree. In contrast, with the filter installed, spurious sound components are effectively suppressed, allowing the true beam pattern of the audio signals generated by SPPAL to be accurately captured. Notably, the beam pattern measurements remain consistent regardless of the incidence angle when the filter is in place.
\par
It should be noted that unnormalized beam patterns are presented here because normalization tends to obscure critical details. Specifically, because the grazing incidence method provides insufficient attenuation, the beam pattern is sharply exaggerated in the paraxial region while accurate measurements are obtained off-axis\textemdash a discrepancy that normalized beam patterns fail to capture. For example, spurious sound is evident in the paraxial region but accurately measured away from the center in \cref{fig:8e}.

\subsubsection{Audio sound propagation curve}
\label{subsubsec:434}

\begin{figure}[t]
  \centering
  \begin{subfigure}[t]{8.5cm}
    \centering
    \includegraphics[width=\textwidth]{Figure/FigureLegend.pdf}
  \end{subfigure}
  \\
  \begin{subfigure}[t]{4.25cm}
    \centering
    \includegraphics[width=\textwidth]{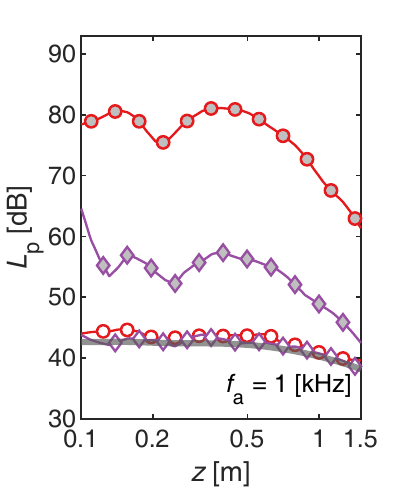}
    \caption{}
    \label{fig:9a}
  \end{subfigure}
  \hfill
  \begin{subfigure}[t]{4.25cm}
    \centering
    \includegraphics[width=\textwidth]{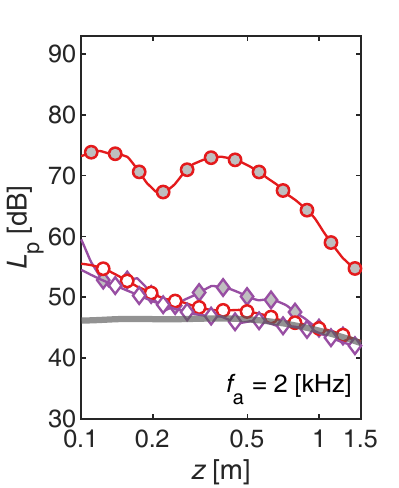}
    \caption{}
    \label{fig:9b}
  \end{subfigure}
  \\
  \begin{subfigure}[t]{4.25cm}
    \centering
    \includegraphics[width=\textwidth]{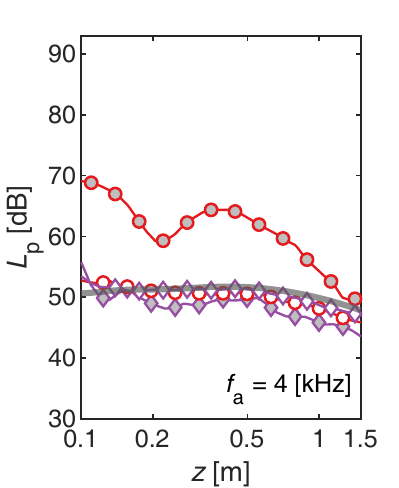}
    \caption{}
    \label{fig:9c}
  \end{subfigure}
  \hfill
  \begin{subfigure}[t]{4.25cm}
    \centering
    \includegraphics[width=\textwidth]{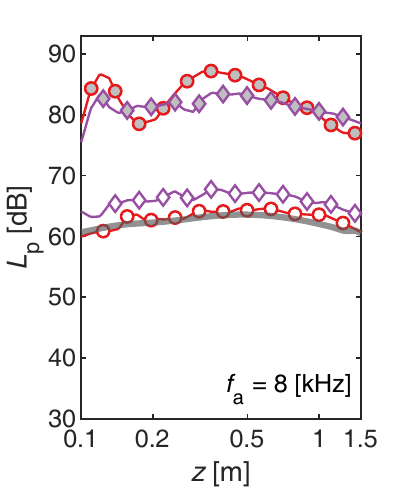}
    \caption{}
    \label{fig:9d}
  \end{subfigure}
  
  \caption{Measured propagation curves of audio signals generated by the SPPAL along the axial direction at audio frequencies of (a) 1 kHz, (b) 2 kHz, (c) 4 kHz, and (d) 8 kHz. Red circles indicate measurements taken at a 0\textdegree\ incidence angle, and purple diamonds represent measurements at a 90\textdegree\ grazing incidence angle. Marker filled color distinguishes between conditions with (white-filled markers) and without (gray-filled markers) the acoustic filter installed. Numerical simulation results obtained from the SWE method are depicted as thick gray solid lines.}
  \label{fig:9}
\end{figure}

\Cref{fig:9} presents the measured propagation curves of audio signals generated by the SPPAL, evaluated along the axial direction at representative audio frequencies of 1, 2, 4, and 8 kHz. The measurements without the acoustic filter reveal significantly elevated sound pressure levels due to prominent spurious sound components, particularly evident at the 0\textdegree\ incidence angle. This artificial elevation hampers the accurate evaluation of the acoustic field of PA.
\par
Conversely, the measurements obtained with the acoustic filter installed demonstrate stabilized and consistent propagation characteristics. The filter effectively suppresses spurious sound, enabling accurate representation of the true audio signal propagation. Additionally, measurements taken with the filter exhibit minimal dependency on incidence angle, confirming that the proposed acoustic filter reliably isolates genuine audio signals from ultrasonic-induced distortions.

\section{Results and Discussion}

The experimental measurements and numerical analysis conducted in this study demonstrate that the proposed half-wavelength acoustic filter effectively addresses the long-standing issue of spurious sound in PA measurements. The measured TL characteristics, frequency responses, beam patterns, and propagation curves consistently validate the acoustic filter's performance, confirming significant suppression of undesired ultrasonic signals.
\par
The TL measurement results, shown in \cref{fig:6}, illustrate extremely high attenuation precisely at the targeted suppression frequencies of 40 kHz and 60 kHz. The experimentally obtained TL closely aligns with FEM simulations, affirming the validity of the simplified, yet robust design methodology employed. Additionally, the measured TL exhibits negligible variability with respect to measurement distance, demonstrating consistent performance throughout typical near-field and far-field scenarios. This distance-independent characteristic is particularly advantageous in practical acoustic measurement setups.
\par
Furthermore, frequency response measurements (\cref{fig:7}) clearly illustrate the filter's effectiveness in eliminating spurious acoustic signals across the audio frequency range. Without the acoustic filter, significant spurious signals distort the measured audio response, particularly at lower frequencies and normal incidence angles. Although grazing incidence angles slightly reduce the measured spurious signals, the suppression remains inadequate. In contrast, with the acoustic filter installed, the audio frequency responses accurately reflect the true audio signals, remaining stable and consistent across different angles and distances. This confirms that the filter's performance is robust against variations in incidence angle, surpassing limitations encountered in conventional grazing incidence methods. Beam pattern measurements, as presented in \cref{fig:8}, further substantiate the filter's importance in accurately evaluating parametric loudspeakers (PALs). Without the acoustic filter, spurious sounds cause exaggerated directivity measurements, falsely indicating higher directional characteristics, particularly at 0\textdegree\ incidence angle. Conversely, the beam patterns measured with the filter installed accurately represent the true directivity of audio signals produced by parametric acoustic arrays. Numerical simulation results reinforce the reliability of the filter-based measurements, facilitating correct characterization of the loudspeaker's acoustic directivity, which is crucial given the fundamental nature of PAL as a directional acoustic source. Additionally, propagation curve measurements (\cref{fig:9}) validate the filter's practical utility in characterizing the spatial distribution of parametric audio fields. Without the acoustic filter, spurious sounds lead to excessively elevated sound pressure levels that distort the propagation curves. The filter installation rectifies this issue, providing stable and realistic propagation curves that closely match the numerical results. Consequently, accurate propagation characterization facilitated by this acoustic filter directly contributes to reliable design, testing, and validation of parametric acoustic loudspeakers.
\par
Compared with previously established methods—such as thin polymer film filters, phononic crystals, and grazing incidence microphones—the proposed acoustic filter demonstrates clear and substantial advantages. It overcomes fabrication difficulties and dimensional constraints encountered with Helmholtz resonators, avoids the incidence angle limitations associated with phononic crystals, and surpasses the modest attenuation achievable by simple angular adjustments of microphones. Moreover, the presented method maintains an exceptionally simple geometry that is readily manufacturable via high-precision SLA 3D printing, facilitating accessibility and reproducibility in various research and industrial contexts.
\par
In conclusion, the proposed half-wavelength resonator acoustic filter significantly enhances measurement accuracy in parametric acoustic array research. By combining the filter with grazing incidence methods, even greater TL can be achieved, enabling accurate measurements in high-intensity ultrasonic fields near the acoustic source. By reliably eliminating spurious signals without complicated apparatus or substantial influence on the acoustic field, it provides researchers and practitioners with a robust and practical tool. This combined approach of filtering and grazing incidence ensures accurate characterization of parametric arrays, making it essential for the precise evaluation and further development of directional acoustic technologies.

\appendix
\counterwithin{figure}{section}
\renewcommand{\thefigure}{\Alph{section}.\arabic{figure}}

\section*{Data availability}
The STL files of the proposed filter for 3D printing are available on \cite{Kim2025}. Caution should be exercised when installing this filter onto the microphone, as blocking the acoustic port may damage the microphone's diaphragm.



\section*{Acknowledgements}
This work was financially supported by the Korea-led K-Sensor Technology Development Program for Market Leadership (RS-2022-00154770) (\textquotedblleft Next-generation piezoelectric based motion sensor platform\textquotedblright) funded by the Ministry of Trade, Industry and Energy (MOTIE, Korea), the POSCO-POSTECH-RIST Convergence Research Center program funded by POSCO, the National Research Foundation (NRF) grants (RS-2024-00356928, RS 2024-00436476) funded by the Ministry of Science and ICT (MSIT) of the Korean government. B. Oh acknowledges the NRF Ph.D. fellowship (RS-2024-00409956) funded by the Ministry of Education of the Korean government.

\section{SPPAL and nonlinear acoustic simulation}
\label{app:1}

\begin{figure}[ht]
    \centering
    \includegraphics[width=0.48\textwidth]{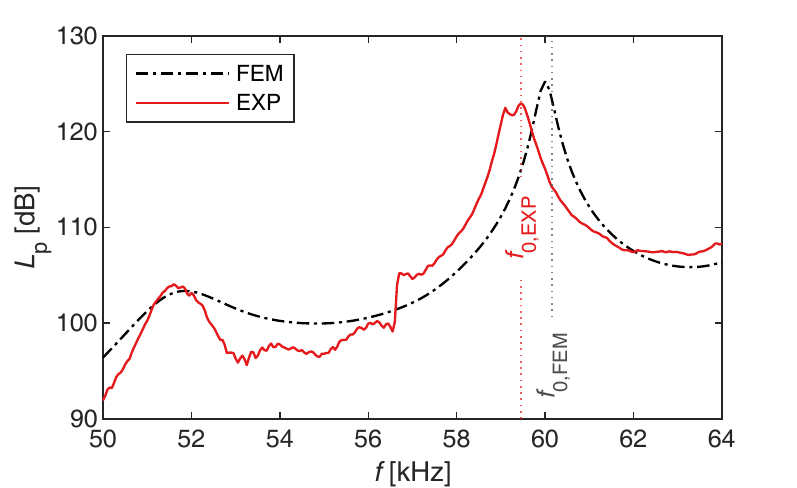}

    \caption{Measured ultrasonic frequency response of the SPPAL (solid line) and the corresponding simulation results (dash-dotted line).}
    \label{fig:a1}
\end{figure}

The SPPAL used in the experimental investigation closely resembles the configuration presented in \cite{Oh2023}. To accurately characterize the transducer for numerical simulations, the ultrasonic frequency response of the fabricated SPPAL was experimentally measured at a distance of $1$~m, as illustrated in \cref{fig:a1}. The measured response served as the basis for adjusting the transducer parameters employed in subsequent FEM simulations. \Cref{fig:a1} compares the measured ultrasonic frequency response with corresponding simulation results, highlighting good agreement yet also indicating minor discrepancies attributable to slight deviations in transducer characteristics.
\par
The nonlinear acoustic analysis of the SPPAL was conducted following a two-step procedure. First, FEM simulations were performed to determine the velocity profile on the radiating surface of the stepped plate. Subsequently, this velocity profile was utilized as the input boundary condition for the nonlinear acoustic field analysis via SWE based on the solution of the Westervelt equation, employing the numerical approach detailed in \cite{Zhong2020,Zhong2024}. This approach enabled accurate prediction of the parametric acoustic field generated by the SPPAL, facilitating comprehensive validation and comparison with the experimental measurements.

\section{Transmission loss and the angle of incidence}
\label{app:2}

\begin{figure}[ht]
    \centering
    \includegraphics[width=0.48\textwidth]{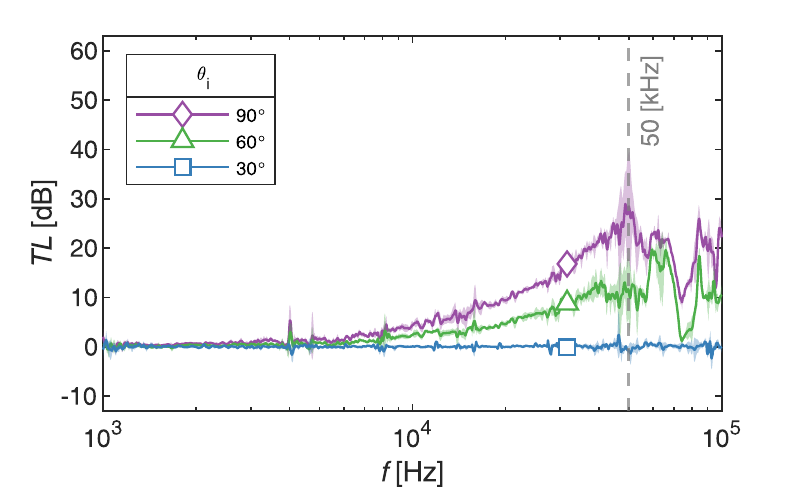}

    \caption{Measured TL for different incidence angles. The solid-colored lines represent TL values averaged from measurements taken at distances of $0.1$, $0.2$, $0.5$, and $1$~m, with red, blue, green, and purple lines corresponding to incidence angles of $30$\textdegree, $60$\textdegree, and $90$\textdegree, with the shaded regions illustrating the standard deviation across distances.}
    \label{fig:b1}
\end{figure}

\Cref{fig:b1} shows the measured TL values for different incidence angles without the acoustic filter. The solid-colored lines represent TL values averaged from measurements taken at distances of $0.1$, $0.2$, $0.5$, and $1$~m, with blue, green, and purple lines corresponding to incidence angles of $30$\textdegree, $60$\textdegree, and $90$\textdegree, respectively. Although adjusting the microphone incidence angle (i.e., grazing incidence method) is a simple approach for spurious sound suppression, the method inherently lacks the ability to precisely control or select the frequency at which maximum TL occurs, limiting its applicability and effectiveness for targeted ultrasonic frequency suppression. In this case, the maximum TL is observed near $50$~kHz.

\bibliographystyle{elsarticle-num} 
\bibliography{ManuscriptNotes.bib}


\end{document}